\documentclass[letter]{jpsj2} 

\usepackage{latexsym}
\usepackage{graphicx}
\usepackage{amsmath}

\def\gs2{0.80} 
\def\g3{1.00} 

\title{Effect of Slow Switching in On-line Learning for Ensemble Teachers}

\author{
\textsc{Seiji MIYOSHI}$^{1}$
\thanks{E-mail address: miyoshi@ipcku.kansai-u.ac.jp}
and \textsc{Masato OKADA}$^{2}$$^{3}$}

\inst{
$^{1}$Department of Electrical and Electronic Engineering, 
Faculty of Engineering Science,\\
 Kansai University, 3-3-35 Yamate-cho, Suita-shi Osaka, 564-8680 \\
$^{2}$Division of Transdisciplinary Sciences, 
Graduate School of Frontier Sciences, \\
The University of Tokyo, 
5--1--5 Kashiwanoha, Kashiwa-shi, Chiba, 277--8561 \\
$^{3}$RIKEN Brain Science Institute, 
2--1 Hirosawa, Wako-shi, Saitama, 351--0198
}

\abst{

We have analyzed the generalization performance of a student
which slowly switches ensemble teachers.
By calculating the generalization error 
analytically using statistical mechanics
in the framework of on-line learning, 
we show that the dynamical behaviors of
generalization error have the periodicity that is 
synchronized with the switching period
and 
the behaviors differ with
the number of ensemble teachers.
Furthermore, we show that 
the smaller the switching period is,
the larger the difference is. 

}

\kword{
ensemble teachers, on-line learning, generalization error,
statistical mechanics, slow switching
}

\begin{document}
\maketitle


Learning can be classified into batch learning and
on-line learning \cite{Saad,Nicolo}.
In on-line learning, examples once used are discarded and
a student cannot give correct answers 
for all examples used in training.
However, there are merits; for example,
a large memory for storing many examples is not necessary
and it is possible to follow a time variant teacher\cite{JPSJ2006,JPSJ2007}. 
%
%
Recently, we used a statistical mechanical method\cite{Saad,NishimoriE}
to analyze the generalization performance
of a model composed of linear perceptrons: 
a true teacher, ensemble teachers, and the student
in the framework of on-line learning\cite{JPSJ2006b}.
That is, we treated a model that has
$K$ teachers called ensemble teachers who 
exist around a true teacher\cite{Hirama}.
%
In the study,
we analyzed the model in which a student switches 
the ensemble teachers in turn or randomly at each time step.
Therefore, the study 
was an analysis of a fast switching model.
On the contrary, 
the properties of a model in which 
a student switches the ensemble teachers slowly
is also attractive.
In this letter, we analyze such a slow switching model.

We have considered a true teacher, $K$ ensemble teachers,
and a student.
They are all linear perceptrons with connection weights
$\mbox{\boldmath $A$}$, 
$\mbox{\boldmath $B$}_k$, and
$\mbox{\boldmath $J$}$, respectively.
Here, $k=1,\ldots,K$.
For simplicity, the connection weight of the true teacher,
the ensemble teachers, and the student
is simply called the true teacher, the ensemble teachers, and
the student, respectively.
The true teacher $\mbox{\boldmath $A$}=\left(A_1,\ldots,A_N\right)$,
ensemble teachers 
$\mbox{\boldmath $B$}_k=\left(B_{k1},\ldots,B_{kN}\right)$,
student
$\mbox{\boldmath $J$}=\left(J_1,\ldots,J_N\right)$,
and input 
$\mbox{\boldmath $x$}=\left(x_1,\ldots,x_N\right)$
are $N$-dimensional vectors.
Each component $A_i$ of $\mbox{\boldmath $A$}$
is drawn from ${\cal N}(0,1)$ independently and fixed,
where ${\cal N}(0,1)$ denotes the Gaussian distribution with
a mean of zero and a variance of unity.
Some components $B_{ki}$
are equal to $A_i$ multiplied by --1,
the others are equal to $A_i$.
Which component $B_{ki}$ is equal to $-A_i$
is independent from the value of $A_i$.
Hence, $B_{ki}$ also obeys ${\cal N}(0,1)$
and it is also fixed.
The direction cosine between 
$\mbox{\boldmath $B$}_k$ and
$\mbox{\boldmath $A$}$ is $R_{Bk}$
and that between
$\mbox{\boldmath $B$}_k$ and
$\mbox{\boldmath $B$}_{k'}$
is $q_{kk'}$.
Each of the components $J_i^0$ 
of the initial value $\mbox{\boldmath $J$}^0$
of $\mbox{\boldmath $J$}$
is drawn from ${\cal N}(0,1)$ independently.
The direction cosine between 
$\mbox{\boldmath $J$}$ and 
$\mbox{\boldmath $A$}$ is  $R_{J}$
and that between
$\mbox{\boldmath $J$}$ and
$\mbox{\boldmath $B$}_{k}$ is $R_{BkJ}$.
Each component $x_i$ of $\mbox{\boldmath $x$}$
is drawn from ${\cal N}(0,1/N)$ independently.
This letter assumes the thermodynamic limit $N\rightarrow \infty$.
Therefore,
$\|\mbox{\boldmath $A$}\|=\|\mbox{\boldmath $B$}_k\|=\|\mbox{\boldmath $J$}^0\|=\sqrt{N}$,
and
$\|\mbox{\boldmath $x$}\|=1$.
Generally, norm $\|\mbox{\boldmath $J$}\|$
of the student
changes as time step proceeds.
Therefore, ratio $l^m$ of the norm to $\sqrt{N}$
is introduced and called the length of 
the student. That is,
$\|\mbox{\boldmath $J$}^m\|=l^m\sqrt{N}$,
where $m$ denotes the time step.
The outputs of the true teacher, the ensemble teachers, 
and the student are
$y^m+n_A^m$, $v_k^m+n_{Bk}^m$ and
$u^ml^m+n_J^m$, respectively.
Here,
$y^m = \mbox{\boldmath $A$}\cdot \mbox{\boldmath $x$}^m$,  
$v_k^m = \mbox{\boldmath $B$}_k\cdot \mbox{\boldmath $x$}^m$, and 
$u^m l^m = \mbox{\boldmath $J$}^m\cdot \mbox{\boldmath $x$}^m$
where
$y^m$, $v_k^m$, and $u^m$
obey Gaussian distributions with a mean of zero and 
a variance of unity.
$n_A^m$, $n_{Bk}^m$, and $n_J^m$ are
independent Gaussian noises with variances of 
$\sigma_{A}^2, \sigma_{Bk}^2$, and $\sigma_J^2$, respectively.

We define the error $\epsilon_{Bk}$ between 
true teacher $\mbox{\boldmath $A$}$
and each member $\mbox{\boldmath $B$}_k$ of 
the ensemble teachers
by the squared errors of their outputs:
$\epsilon_{Bk}^m \equiv \frac{1}{2}\left( y^m+n_{A}^m - v_k^m -n_{Bk}^m\right)^2$.
%
In the same manner,
we define error $\epsilon_{BkJ}$ between 
each member $\mbox{\boldmath $B$}_k$ of 
the ensemble teachers
and student $\mbox{\boldmath $J$}$
by the squared errors of their outputs:
$\epsilon_{BkJ}^m \equiv \frac{1}{2}\left( v_k^m+n_{Bk}^m-u^ml^m-n_J^m\right)^2$.
Student $\mbox{\boldmath $J$}$ 
adopts the gradient method as a learning rule
and uses
input $\mbox{\boldmath $x$}$
and an output
of one of the $K$ ensemble teachers
$\mbox{\boldmath $B$}_k$.
Here, the student $\mbox{\boldmath $J$}$
uses each ensemble teacher $\mbox{\boldmath $B$}_k$
$TN$ times succsessively
where $T$ is $O(1)$.
That is,
\begin{eqnarray}
\mbox{\boldmath $J$}^{m+1}
&=& \mbox{\boldmath $J$}^{m} 
   -\eta \frac{\partial \epsilon_{BkJ}^m}{\partial \mbox{\boldmath $J$}^{m}}\\
&=&  \mbox{\boldmath $J$}^{m} 
   +\eta \left( v_k^m+n_{Bk}^m-u^ml^m-n_J^m\right)
   \mbox{\boldmath $x$}^{m},\label{eqn:Jupdate}\\
k&=& \mbox{mod}\left(\left[\frac{m}{TN}\right],K\right)+1,\label{eqn:SW}
\end{eqnarray}
where $\eta$ denotes the learning rate
and is a constant number.
The Gauss notation
is denoted by $\left[\cdot\right]$.
That is,
$\left[\frac{m}{TN}\right]$
is the maximum integer which is not larger than $\frac{m}{TN}$.
Here, 
$\mbox{mod}\left(\left[\frac{m}{TN}\right],K\right)$
denotes
the remainder of $\left[\frac{m}{TN}\right]$ divided by $K$.
Equation (\ref{eqn:SW}) means that 
the student 
uses each ensemble teacher
$TN \sim O(N)$ times succsessively. We call this 
slow switching.
By generalizing the learning rules, 
Eq. (\ref{eqn:Jupdate})
can be expressed as
$\mbox{\boldmath $J$}^{m+1}
= \mbox{\boldmath $J$}^{m}
 +f_k
 \mbox{\boldmath $x$}^{m}$, 
where $f$ denotes a function
that represents the update amount and 
is determined by the learning rule.
In addition, we define the error $\epsilon_J$ between 
true teacher $\mbox{\boldmath $A$}$ and 
student $\mbox{\boldmath $J$}$ by 
the squared error of their outputs:
$\epsilon_J^m \equiv \frac{1}{2}\left( y^m+n_{A}^m -u^ml^m-n_J^m\right)^2$.

One of the goals of statistical learning theory
is to theoretically obtain generalization errors.
Since generalization error is the mean of errors 
for the true teacher 
over the distribution of new input and noises,
generalization error $\epsilon_{Bkg}$ of 
each member $\mbox{\boldmath $B$}_k$ of the ensemble teachers
and $\epsilon_{Jg}$ of student
$\mbox{\boldmath $J$}$
are calculated as follows.
Superscripts $m$, which represent the time step, 
are omitted for simplicity unless stated otherwise.
\begin{eqnarray}
\epsilon_{Bkg}
&=& \int d\mbox{\boldmath $x$} dn_{A} dn_{Bk} 
         P\left(\mbox{\boldmath $x$}, n_{A}, n_{Bk}\right)
         \epsilon_{Bk}
         \\
&=& \int dy dv_k  dn_{A} dn_{Bk} 
    P\left(y, v_k, n_{A}, n_{Bk}\right)
    \frac{1}{2}
    \left( y+n_{A} - v_k -n_{Bk}\right)^2 \\
&=& \frac{1}{2}
    \left(
    -2R_{Bk} + 2 + \sigma_{A}^2 + \sigma_{Bk}^2
    \right),
    \label{eqn:hateg} \\
\epsilon_{Jg} 
&=& \int d\mbox{\boldmath $x$} dn_{A} dn_J 
         P\left(\mbox{\boldmath $x$}, n_{A}, n_J\right)
         \epsilon_J
         \\
&=& \int dy du dn_{A} dn_J 
    P\left(y, u, n_{A}, n_J\right)
    \frac{1}{2}
    \left(y+n_{A} - ul - n_J \right)^2 \\
&=& \frac{1}{2}
    \left(
    -2l R_J + l^2 + 1 + \sigma_{A}^2 + \sigma_J^2
    \right).
    \label{eqn:eJg}
\end{eqnarray}


To simplify the analysis, two auxiliary order parameters
$r_J\equiv l R_J$ and $r_{BkJ}\equiv l R_{BkJ}$
are introduced.
Simultaneous differential equations
in deterministic forms \cite{NishimoriE}, 
which describe the dynamical behaviors of order parameters
when the student uses a teacher $\mbox{\boldmath $B$}_{k'}$
that consists of ensemble teachers
have been obtained on the basis of self-averaging
in the thermodynamic limits as follows:
\begin{align}
\frac{dr_{BkJ}}{dt}&= \langle f_{k'} v_k\rangle,& 
\frac{dr_J}{dt}&= \langle f_{k'} y\rangle,& 
\frac{dl}{dt}&= \langle f_{k'} u\rangle+\frac{1}{2l}\langle f_{k'}^2 \rangle.
\label{eqn:dldt}
\end{align}
Here, dimension $N$ has been treated 
to be sufficiently greater than
the number $K$ of ensemble teachers.
Time is defined by $t=m/N$, that is, 
time step $m$ normalized by dimension $N$.
Since linear perceptrons are treated in this letter,
the sample averages that appeared in the above 
equations can be easily calculated as follows:
\begin{align}
\langle f_{k'} u \rangle &= \eta \left(\frac{r_{Bk'J}}{l}-l\right), &
\langle f_{k'}^2 \rangle &= 
\eta^2 \left(l^2-2r_{Bk'J}+1+\sigma_{Bk'}^2+\sigma_J^2\right), \\
\langle f_{k'} y \rangle &= \eta \left(R_{Bk'} -r_J\right), &
\langle f_{k'} v_k \rangle
&= \eta \left(q_{k'k}-r_{BkJ}\right). \label{eqn:fk'}
\end{align}

Let us denote the values of 
$r_J, r_{BkJ}$, and $l^2$ of $t=t_0$
as 
$r_J^{t_0}, r_{BkJ}^{t_0}$, and $(l^2)^{t_0}$, respectively.
By using these as intitial values,
simultaneous differential equations
Eqs.(\ref{eqn:dldt})--(\ref{eqn:fk'})
can be solved analytically as follows:
\begin{eqnarray}
r_{BkJ}&=& q_{k'k}+\left(r_{BkJ}^{t_0}-q_{k'k}\right)e^{-\eta (t-t_0)}, 
\label{eqn:rBkJ} \\
r_J &=& R_{Bk'}+\left(r_J^{t_0}-R_{Bk'}\right)e^{-\eta(t-t_0)},
\label{eqn:rJ} \\
l^2 &=& 
1+\frac{\eta}{2-\eta}\left(\sigma_{Bk'}^2+\sigma_J^2\right)
+2\left(r_{Bk'J}^{t_0}-1\right)e^{-\eta(t-t_0)} \nonumber \\
& & +\left(
(l^2)^{t_0}-1-\frac{\eta}{2-\eta}\left(\sigma_{Bk'}^2+\sigma_J^2\right)
-2\left(r_{Bk'J}^{t_0}-1\right)\right)e^{\eta(\eta-2)(t-t_0)}
.\label{eqn:l2} 
\end{eqnarray}

Since all components $A_i$ and $J_i^0$ 
of true teacher $\mbox{\boldmath $A$}$
and the initial student $\mbox{\boldmath $J$}^0$
are drawn from ${\cal N}(0,1)$ independently,
and because the thermodynamic limit $N\rightarrow \infty$
is also assumed,
they are orthogonal to each other at $t=0$.
That is, $R_J=0$ and $l=1$ when $t=0$.

%

In the following,
we consider the case where direction cosines $R_{Bk}$ between 
the ensemble teachers and the true teacher,
direction cosines $q_{kk'}$ among the ensemble teachers
and variances $\sigma_{Bk}^2$ of the noises of 
ensemble teachers are uniform.
That is,
\begin{equation}
R_{Bk} = R_B, (k=1,\ldots,K),\ \ \ \ 
q_{kk'}
=\left\{
\begin{array}{ll}
1,           & (k=k') , \\
q,           & (\text{otherwise}),
\end{array}
\right.\ \ \ \ 
\sigma_{Bk}^2  = \sigma_B^2.
\end{equation}

\begin{figure}[!htbp] 
\begin{minipage}{.50\linewidth}
\begin{flushright}
\includegraphics[width=\gs2\linewidth,keepaspectratio]{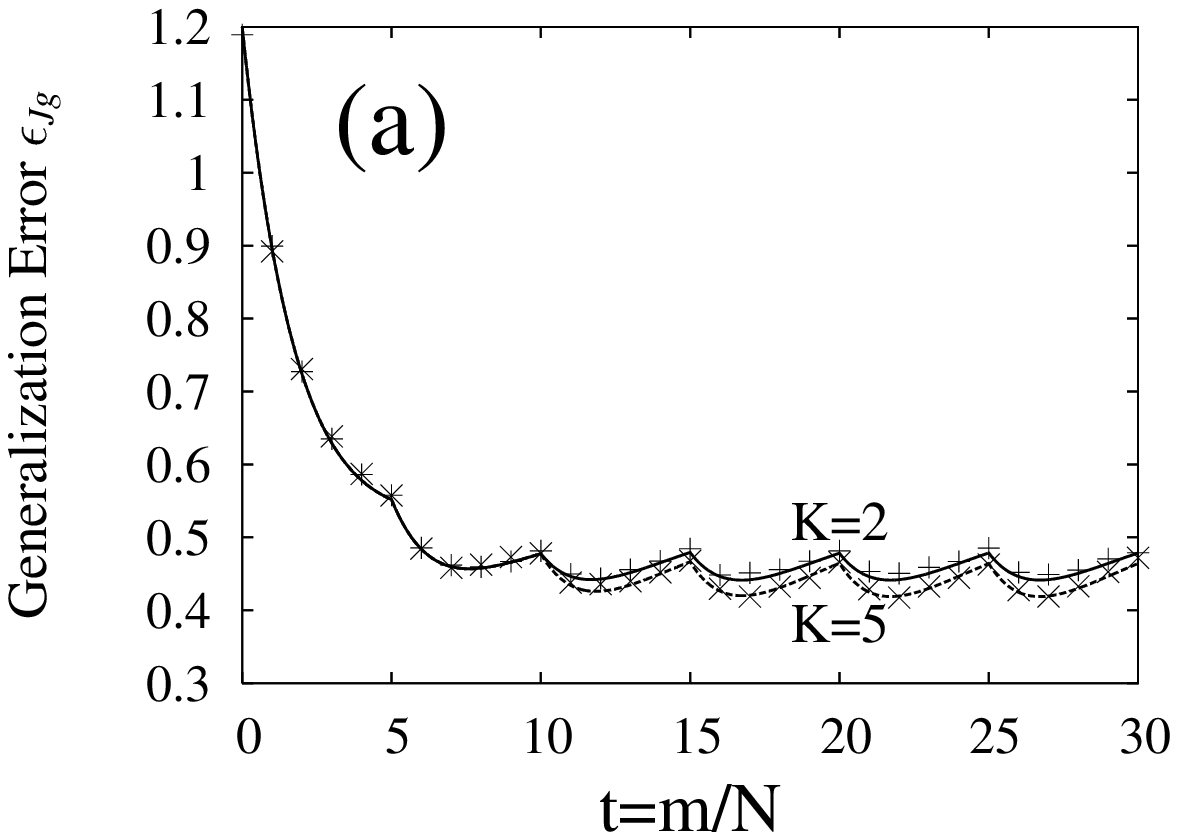}
\end{flushright}
\end{minipage}
\begin{minipage}{.50\linewidth}
\begin{flushleft}
\includegraphics[width=\gs2\linewidth,keepaspectratio]{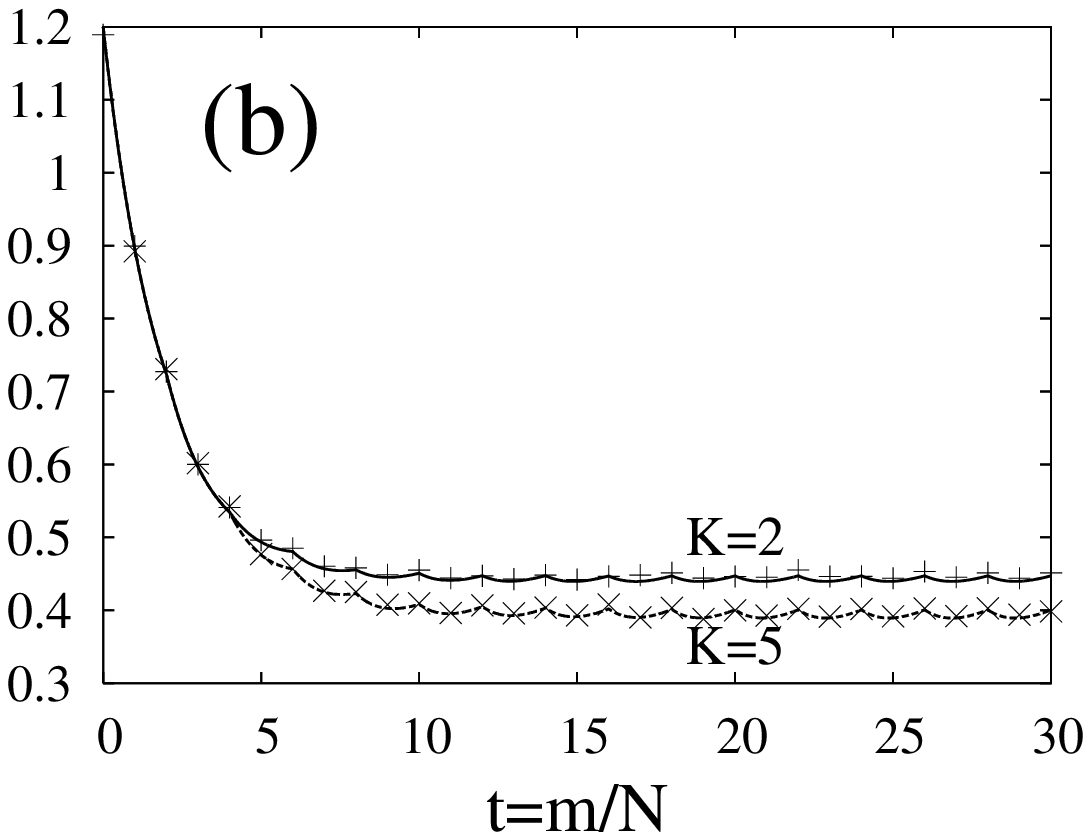}
\end{flushleft}
\end{minipage}
\caption{Dynamical behaviors of generalization errors 
$\epsilon_{Jg}$ when $\eta=0.3$. 
Theory and computer simulations.
$R_B=0.7, q=0.49, \sigma_A^2=0.1, 
\sigma_B^2=0.2$, and $\sigma_J^2=0.3$.
(a)$T=5.0$,
(b)$T=2.0$.
}
\label{fig:egRB07q049E03VA010VB020VJ030}
\end{figure}

\begin{figure}[!htbp]
\begin{minipage}{.50\linewidth}
\begin{flushright}
\includegraphics[width=\gs2\linewidth,keepaspectratio]{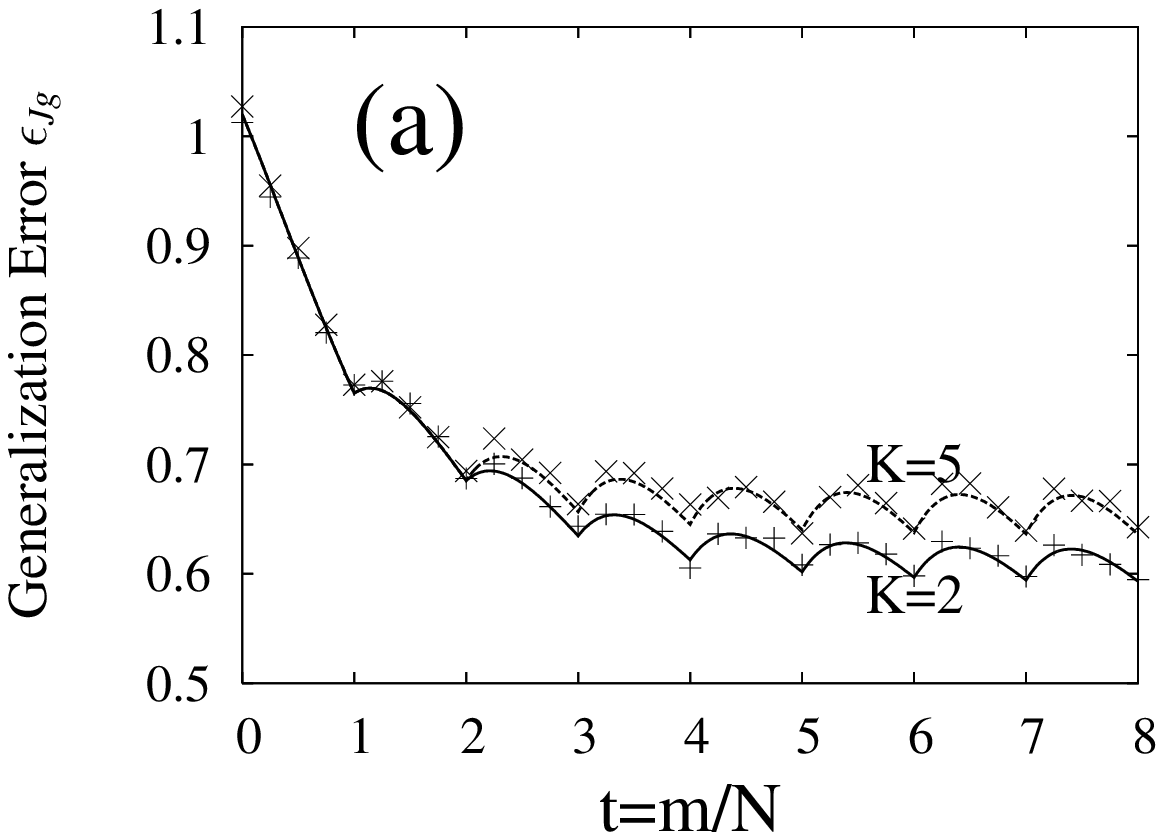}
\end{flushright}
\end{minipage}
\begin{minipage}{.50\linewidth}
\begin{flushleft}
\includegraphics[width=\gs2\linewidth,keepaspectratio]{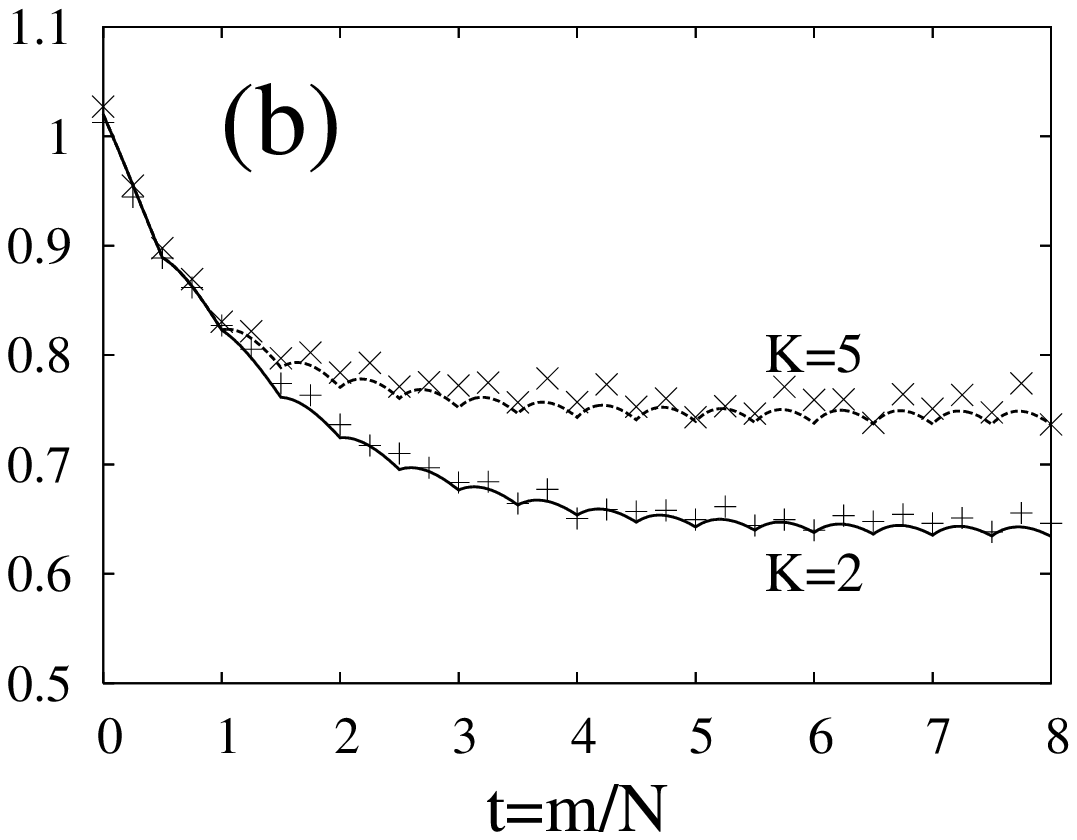}
\end{flushleft}
\end{minipage}
\caption{Dynamical behaviors of generalization errors 
$\epsilon_{Jg}$ when $\eta=1.5$. 
Theory and computer simulations.
$R_B=0.7, q=0.49, \sigma_A^2=0.01, 
\sigma_B^2=0.02$, and $\sigma_J^2=0.03$.
(a)$T=1.0$, (b)$T=0.5$.}
\label{fig:egRB07q049E15VA001VB002VJ003}
\end{figure}

The dynamical behaviors of generalization errors
$\epsilon_{Jg}$
have been analytically obtained by substituting 
Eqs. (\ref{eqn:rJ}) and (\ref{eqn:l2}) into
Eq. (\ref{eqn:eJg}).
The analytical results and the corresponding 
simulation results, where $N=10^5$
are shown in
Figs. \ref{fig:egRB07q049E03VA010VB020VJ030}
and
\ref{fig:egRB07q049E15VA001VB002VJ003}.
In computer simulations, 
$\epsilon_{Jg}$
was obtained by
averaging the squared errors for $5\times10^4$ random inputs 
at each time step.
In these figures, 
the curves represent theoretical results. 
The symbols 
represent simulation results.
In these figures, 
$R_B=0.7$ and $q=0.49$ are common conditions.
In addition, 
$\eta=0.3, \sigma_A^2=0.1, \sigma_B^2=0.2$, and $\sigma_J^2=0.3$
are conditions for
Fig. \ref{fig:egRB07q049E03VA010VB020VJ030}.
$\eta=1.5, \sigma_A^2=0.01, \sigma_B^2=0.02$, and $\sigma_J^2=0.03$
are conditions for
Fig. \ref{fig:egRB07q049E15VA001VB002VJ003}.


These figures show
that the dynamical behaviors of
generalization error have the periodicity that is 
synchronized with the switching period $T$.
In the case of $K=2$,
the student uses ensemble teachers
as
$
\mbox{\boldmath $B$}_1 \rightarrow
\mbox{\boldmath $B$}_2 \rightarrow
\mbox{\boldmath $B$}_1 \rightarrow
\mbox{\boldmath $B$}_2 \rightarrow
\cdots$
. In the case of $K=5$,
$
\mbox{\boldmath $B$}_1 \rightarrow
\mbox{\boldmath $B$}_2 \rightarrow
\mbox{\boldmath $B$}_3 \rightarrow
\mbox{\boldmath $B$}_4 \rightarrow
\mbox{\boldmath $B$}_5 \rightarrow
\mbox{\boldmath $B$}_1 \rightarrow
\mbox{\boldmath $B$}_2 \rightarrow
\mbox{\boldmath $B$}_3 \rightarrow
\cdots$.
Therefore, by comparing the behaviors of $K=2$
and that of $K=5$,
the generarization errors $\epsilon_{Jg}$ completely agree
during the time corresponding to two cycles from the initial state
because the teachers used by student are the same.
On the contrary, 
the generarization errors $\epsilon_{Jg}$ 
of $K=2$ and $K=5$ are not the same
after the second cycle.
In our study on the fast switching model\cite{JPSJ2006b},
it was proven that
when a student's learning rate satisfies $\eta <1$,
the larger the number $K$ is,
the smaller the student's generalization error is.
The same phenomenon is observed in the slow switching model
treatd in this letter, that is,
the generalization error of $K=5$ is smaller than that of $K=2$
as shown in Fig. \ref{fig:egRB07q049E03VA010VB020VJ030}.
On the contrary,
the generalization error of $K=5$ is larger than that of $K=2$
in Fig. \ref{fig:egRB07q049E15VA001VB002VJ003}.
Here, the dynamical behavior approaches that of 
the fast switching model\cite{JPSJ2006b} asymptotically
in the limit of switching period $T\rightarrow 0$.


In both cases of $\eta=0.3$ and $1.5$,
the smaller the switching period $T$ is,
the larger the difference between the
generalization error $\epsilon_{Jg}$ of $K=2$
and that of $K=5$ is.
The reason is the following:
if the switching period $T$ is large,
a student learns enough from only 
the one teacher that the student uses in the period.
In other words, as the student forgets the other teachers,
the influence of the number $K$ of ensemble 
teachers becomes small.


\begin{figure}[!htbp]
\begin{minipage}{.50\linewidth}
\begin{flushright}
\includegraphics[width=0.80\linewidth,keepaspectratio]{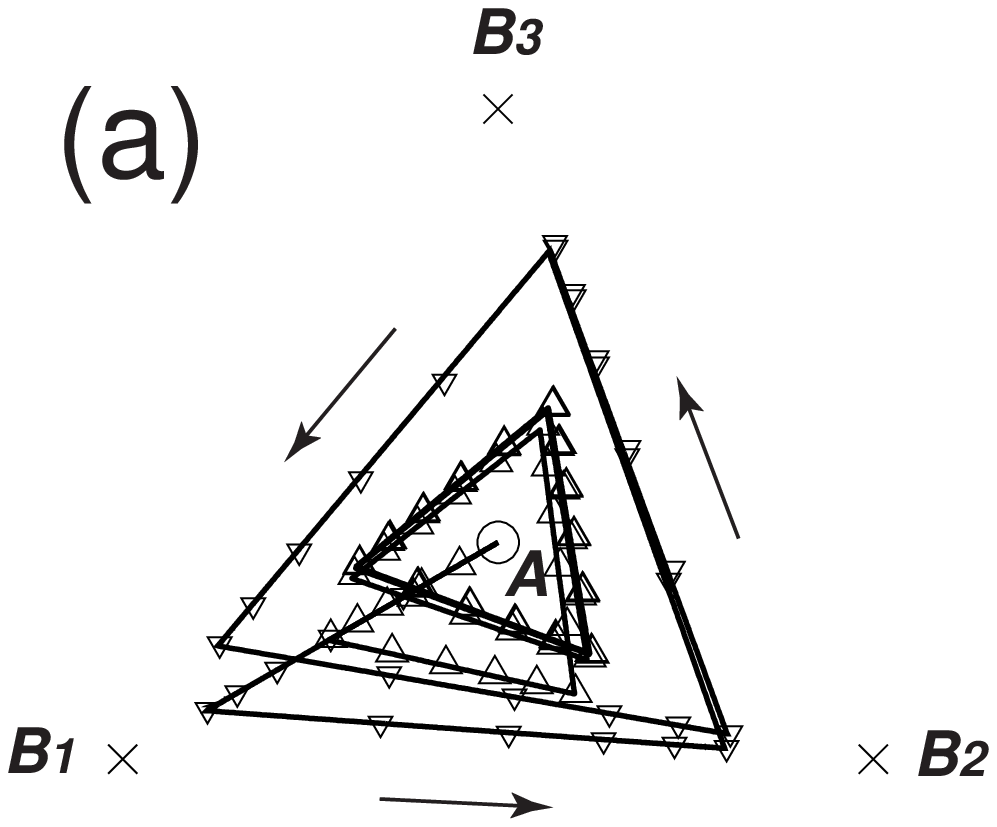}
\end{flushright}
\end{minipage}
\begin{minipage}{.50\linewidth}
\begin{flushleft}
\includegraphics[width=0.80\linewidth,keepaspectratio]{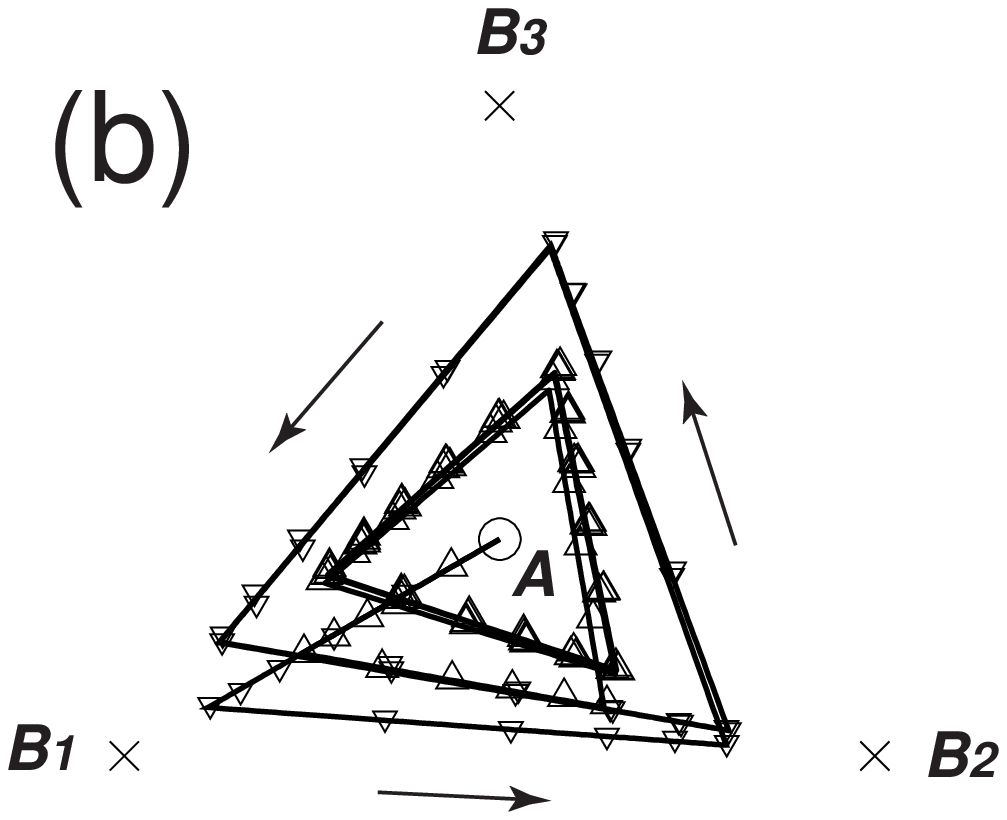}
\end{flushleft}
\end{minipage}
\caption{Student's projection to 
2-D plane on which $\mbox{\boldmath $B$}_1$--$\mbox{\boldmath $B$}_3$ exist.
(a)$\eta=0.3$, (b)$\eta=1.5$.
Solid lines
represent 
trajectories of student's projection obtained theoretically.
Symbols $\bigtriangleup$ and 
$\bigtriangledown$ 
represent computer simulations with 
(a)$T=2.0$ and $T=5.0$, 
(b)$T=0.5$ and $T=1.0$, respectively.
}
\label{fig:EJ}
\end{figure}

We visualize the student's behaviors in the case of $K=3$ 
to understand them intuitively.
That means we obtain the student's projection to 
the two-dimensional plane on which the three ensemble teachers exist.
Figure \ref{fig:EJ}
shows the projection's trajectories in the case of 
$\eta=0.3$ and $\eta=1.5$.
In this figure, 
symbols $\times$, 
$\circ$ and 
solid lines
represent 
the ensemble teachers $\mbox{\boldmath $B$}_1$, $\mbox{\boldmath $B$}_2$ and $\mbox{\boldmath $B$}_3$,
the projection of the true teacher $\mbox{\boldmath $A$}$ and
the trajectories of the student's projection obtained theoretically, respectively. 
In Fig. \ref{fig:EJ}(a), 
symbols $\bigtriangleup$ and 
$\bigtriangledown$ 
represent 
the student's projections obtained by computer simulations with 
$T=2.0$ and $T=5.0$, respectively.
In Fig. \ref{fig:EJ}(b), 
those represent 
the projections with 
$T=0.5$ and $T=1.0$, respectively.
This figure shows that
the student moves straight toward the teacher that the student uses then.
Therefore, the student's trajectories in the steady state are regular triangles.
The triangles are small when the switching period $T$ is small
and the triangles are large when $T$ is large. 
In this figure, a side of the trajectory corresponds to
a period in 
Figs. \ref{fig:egRB07q049E03VA010VB020VJ030}
and
\ref{fig:egRB07q049E15VA001VB002VJ003}.
Note that the distance between the student and the true teacher
in Fig. \ref{fig:EJ}
is not necessarily related to the real distance 
between the student and the true teacher
nor the generalization error
since this figure shows the projections.
Though the student is near the true teacher
when $T$ is small in Fig. \ref{fig:EJ}(b),
the generalization error is small when $T$ is large
as shown in 
Fig. \ref{fig:egRB07q049E15VA001VB002VJ003}.


\begin{figure}[!htbp]
\begin{minipage}{.50\linewidth}
\begin{flushright}
\includegraphics[width=\gs2\linewidth,keepaspectratio]{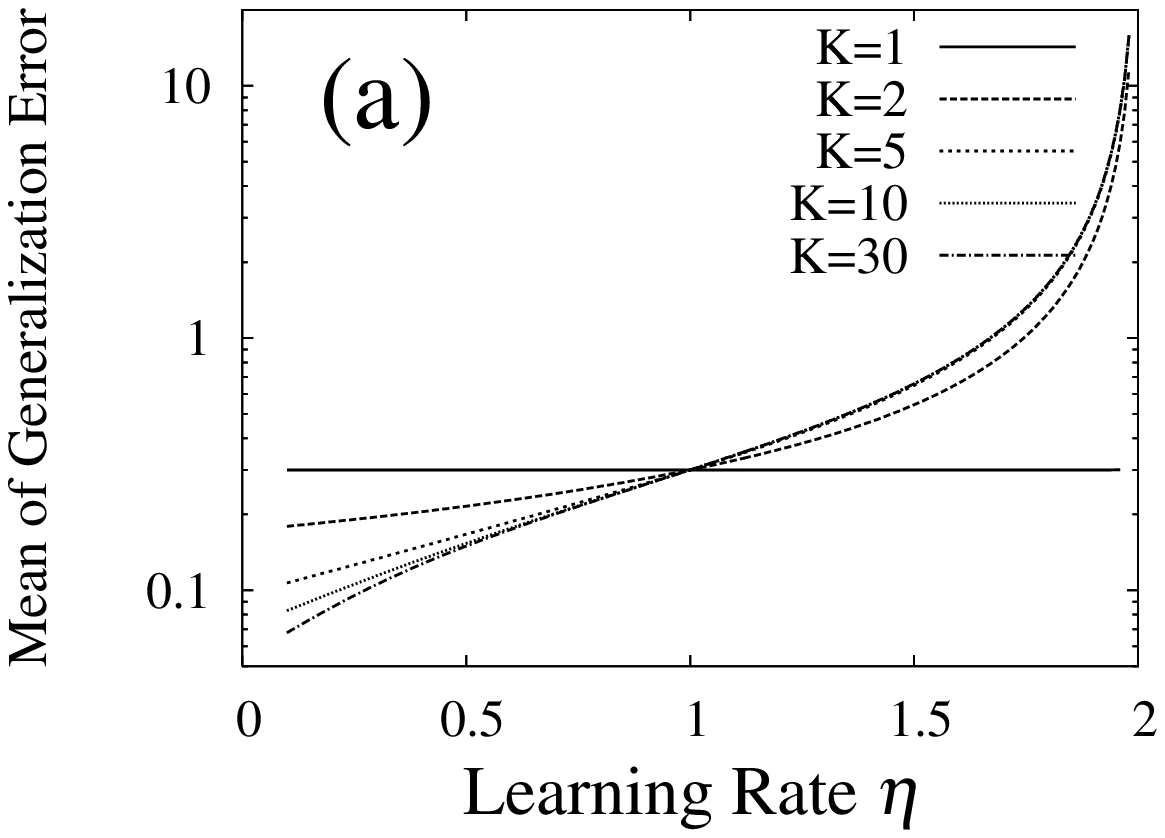}
\end{flushright}
\end{minipage}
\begin{minipage}{.50\linewidth}
\begin{flushleft}
\includegraphics[width=\gs2\linewidth,keepaspectratio]{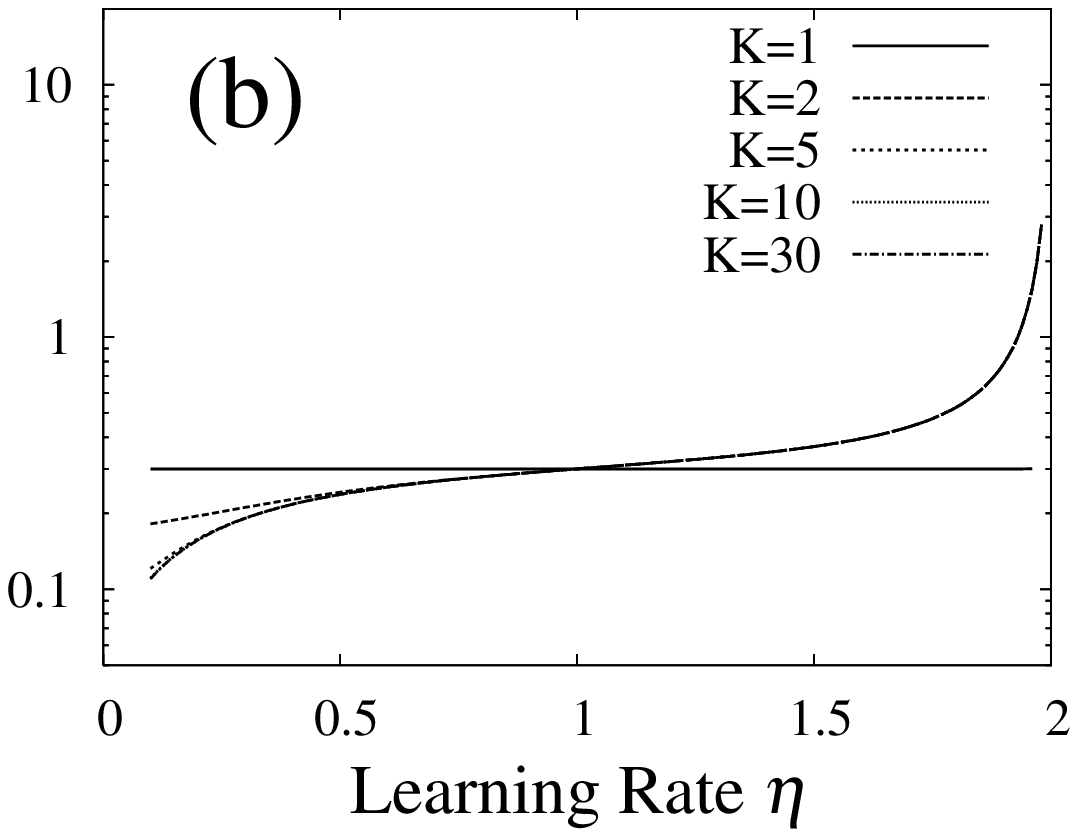}
\end{flushleft}
\end{minipage}
\caption{Means of steady state generalization errors 
$\epsilon_{Jg}$. Theory.
$q=0.49, R_B=0.7$ and $\sigma_A^2=\sigma_B^2=\sigma_J^2=0.0$.
(a)$T=0.5$, (b)$T=5.0$.
}
\label{fig:GR070q049VA00VB00VJ00}
\end{figure}

The relationships between the learning rate $\eta$ and the means of 
steady state generalization errors $\epsilon_{Jg}$
are shown in
Fig. \ref{fig:GR070q049VA00VB00VJ00}.
The means are measured by averaging the generalization errors
during a cycle after the dynamical behaviors reach 
the steady state.
In this figure,
when a learning rate satisfies $\eta <1$,
the larger the number $K$ is,
the smaller the generalization error is.
This is the same property with that of 
the fast switching model\cite{JPSJ2006b}.
A comparison of 
Figs. \ref{fig:GR070q049VA00VB00VJ00}(a) and
\ref{fig:GR070q049VA00VB00VJ00}(b)
shows that
the smaller the switching period $T$ is,
the larger the difference among the
means of generalization errors $\epsilon_{Jg}$ of various $K$
values in the slow switching model as treated in this letter.


\begin{figure}[htbp]
\begin{minipage}{.50\linewidth}
\begin{flushright}
\includegraphics[width=\gs2\linewidth,keepaspectratio]{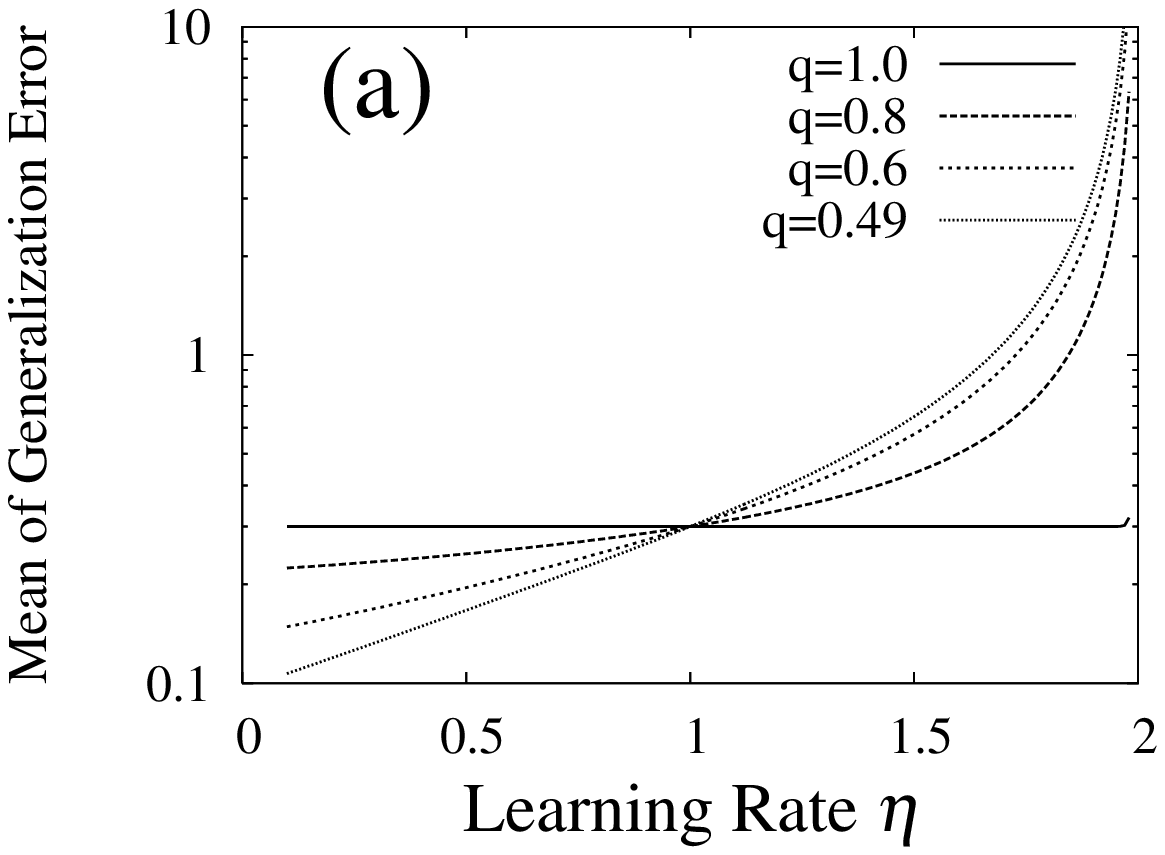}
\end{flushright}
\end{minipage}
\begin{minipage}{.50\linewidth}
\begin{flushleft}
\includegraphics[width=\gs2\linewidth,keepaspectratio]{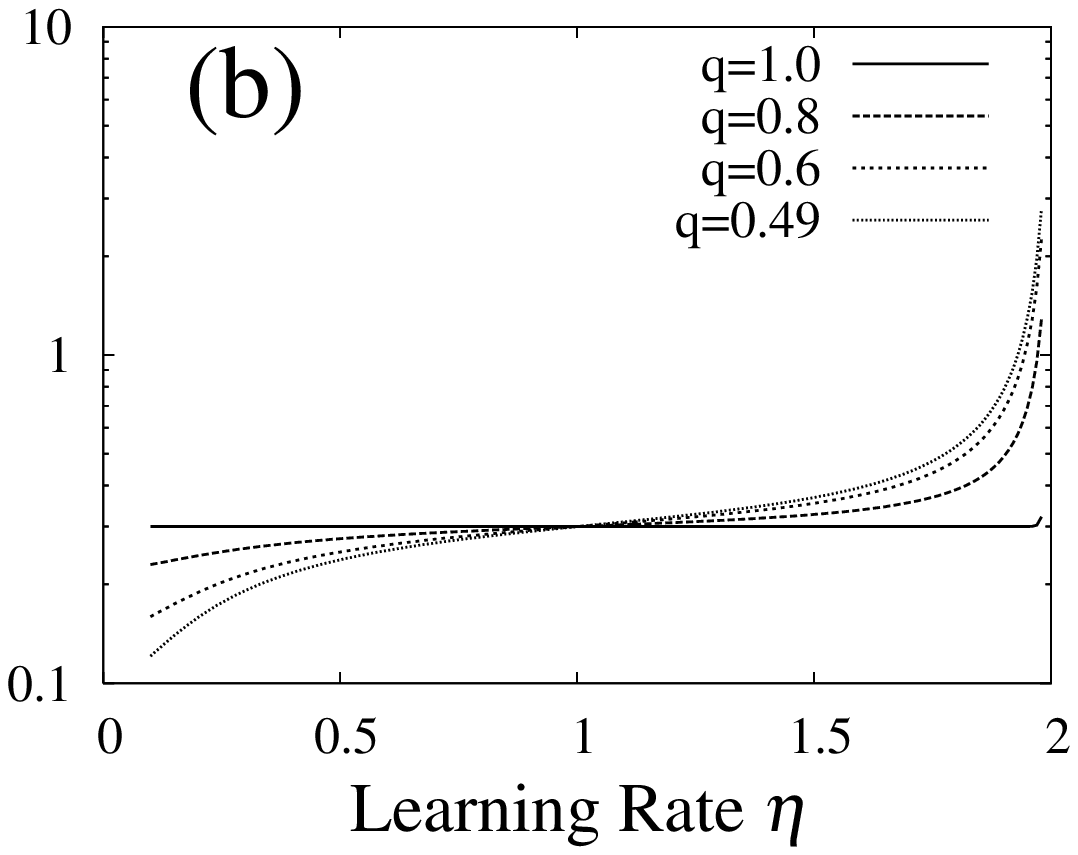}
\end{flushleft}
\end{minipage}
\caption{Means of steady state generalization errors 
$\epsilon_{Jg}$. Theory.
$K=5, R_B=0.7$ and $\sigma_A^2=\sigma_B^2=\sigma_J^2=0.0$.
(a)$T=0.5$, (b)$T=5.0$.
}
\label{fig:GR070K5VA00VB00VJ00}
\end{figure}

The relationships between the learning rate $\eta$ and the means of
steady state generalization errors $\epsilon_{Jg}$
for various direction cosines $q$ are shown in
Fig. \ref{fig:GR070K5VA00VB00VJ00}.
As shown in this figure,
when a learning rate satisfies $\eta <1$,
the smaller $q$ is,
the smaller the generalization error is.
This is also the same property as that of 
the fast switching model\cite{JPSJ2006b}.
By comparing
Figs. \ref{fig:GR070K5VA00VB00VJ00}(a) and
\ref{fig:GR070K5VA00VB00VJ00}(b),
we see that
the smaller the switching period $T$ is,
the larger the difference among the
means of generalization errors $\epsilon_{Jg}$ of various $q$.

In summary, we have analyzed the generalization performance of a student 
in a model composed of linear perceptrons: 
a true teacher, ensemble teachers, and the student. 
In particular, the case where the student slowly switches
ensemble teachers has been analyzed.
By calculating the generalization error 
analytically using statistical mechanics
in the framework of on-line learning, 
we have shown that the dynamical behaviors of
generalization error have the periodicity that is 
synchronized with the switching period
and that 
the behaviors differ with
the number of ensemble teachers.
Furthermore, we have shown that 
the smaller the switching period is,
the larger the difference is.

\section*{Acknowledgments}
This research was partially supported by the Ministry of Education, 
Culture, Sports, Science, and Technology of Japan, 
with Grants-in-Aid for Scientific Research
16500093, 18020007, 18079003, and 18500183.
%



\begin{thebibliography}{99}


\bibitem{Saad}
D.~Saad, (ed.):
{\it On-line Learning in Neural Networks}
(Cambridge University Press, Cambridge, 1998).

\bibitem{Nicolo}
N.~Cesa-Bianchi and G.~Lugosi:
{\it Prediction, Learning, and Games}
(Cambridge University Press, New York, 2006).



\bibitem{JPSJ2006}
S.~Miyoshi and M.~Okada:
J. Phys. Soc. Jpn. {\bf 75} (2005) 024003.

\bibitem{JPSJ2007}
M.~Urakami, S.~Miyoshi, and M.~Okada:
J. Phys. Soc. Jpn. {\bf 76} (2005) 044003.











\bibitem{NishimoriE}
H.~Nishimori:
{\it Statistical Physics of Spin Glasses and Information Processing: 
An Introduction}
(Oxford University Press, Oxford, 2001).


\bibitem{JPSJ2006b}
S.~Miyoshi and M.~Okada:
J. Phys. Soc. Jpn. {\bf 75} (2006) 044002.

\bibitem{Hirama}
T.~Hirama and K.~Hukushima:
J. Phys. Soc. Jpn. {\bf 77} (2008) 094801.


\end{thebibliography}
\end{document}